\documentclass[aps,prl,superscriptaddress,twocolumn,amsfonts,amsmath,a4paper,showpacs]{revtex4}
\usepackage{graphicx}

\begin{document}

\title{Entangled-like Chain Dynamics in Non-entangled Polymer Blends \\ with Large Dynamic Asymmetry}

\author{Angel J. Moreno}
\affiliation{Centro de F\'{\i}sica de Materiales (CSIC-UPV/EHU), 
Apartado 1072, 20080 San Sebasti\'{a}n, Spain.}
\author{Juan Colmenero}
\affiliation{Centro de F\'{\i}sica de Materiales (CSIC-UPV/EHU), 
Apartado 1072, 20080 San Sebasti\'{a}n, Spain.}
\affiliation{\mbox{Departamento de F\'{\i}sica de Materiales, Universidad del Pa\'{\i}s Vasco (UPV/EHU),
Apartado 1072, 20080 San Sebasti\'{a}n, Spain.}}
\affiliation{Donostia International Physics Center, Paseo Manuel de Lardizabal 4,
20018 San Sebasti\'{a}n, Spain.}

\begin{abstract}

We discuss simulations of a simple model for polymer blends in the framework of the Rouse model. 
At odds with standard predictions, large dynamic asymmetry between the two components
induces strong non-exponentiality of the Rouse modes for the fast component. 
Despite chains being much shorter than the entanglement length, it also induces dynamic features
resembling a crossover to entangled-like chain dynamics. This unusual behavior
is associated to strong memory effects which break down the assumption
of time uncorrelation of the external forces acting on the tagged chain.

\end{abstract}

\date{\today}
\maketitle

Polymer blends are systems of wide technological interest whose rheological properties
can be tuned by varying the mixture composition.
They are dynamically heterogeneous: starting from two homopolymers with different mobilities,
two separated segmental relaxations are still observed in the blend state.
In the usual scenario the two components display qualitatively similar dynamic features in the blend. 
A rather different scenario arises, for low concentration of the fast component,
if the two homopolymers exhibit very different glass transition temperatures \cite{nmr,genix}. 
In this case the two components exhibit strong dynamic inmiscibility,
with a large separation (dynamic asymmetry) in their respective relaxation times. The latter can be of even 12 decades 
for high dilution of poly(ethylene oxide) (PEO) in
poly(methyl methacrylate) (PMMA) \cite{nmr}. In such conditions the motion of the
fast component seems to be strongly confined by the slowly relaxing matrix formed by the slow component.
We have recently performed a computational investigation of the structural $\alpha$-relaxation
in a simple bead-spring model for polymer blends \cite{blendpaper}. 
The introduction of size disparity between
the monomers of the two components induces a large time scale separation
at low concentrations of the fast component, which displays unusual
dynamic features, as logarithmic relaxation for density correlators,
or dynamic decoupling between self-motions and intrachain collective relaxation.
These unusual observations are supported by recent
fully atomistic simulations \cite{genix} and neutron scattering experiments \cite{niedzwiedz}
on the real blend PEO/PMMA.

In this Letter we show that anomalous dynamic features induced by confinement on the fast component
extend to large scales. By analyzing relaxation of the chain degrees of freedom
within the framework of the Rouse model \cite{doibook}
novel features are observed for the behavior of the chain normal modes of the fast component.
We simulate short chains ($4\le N \le 21$ monomers) corresponding to the non-entangled case in the homopolymer state 
(the entanglement length \cite{kremer} is $N_{\rm e} \sim 35$ monomers). 
This choice allows us to compare results between the blend
and homopolymer states for chain lengths for which the homopolymer case shows reasonable
agreement with predictions of the Rouse model \cite{kremer,bennemannrouse,shaffer,kreer}.
We test the two basic predictions: orthogonality and exponentiality of the Rouse modes.
Simulations show that orthogonality is a reasonable approximation for both components, 
as well as exponentiality in the case of the slow one. However, large deviations from exponentiality
are found for the fast component in the blend.
We observe a striking crossover, by increasing the dynamic asymmetry in the blend,
to a regime resembling scaling features characteristic of entangled-like chain dynamics,
despite the used chain length being much shorter than $N_{\rm e}$. 
Anomalous scaling is found even in the short-chain limit.
We associate the mentioned anomalous features for the Rouse modes of the fast component
to strong memory effects, which are induced by the slow nature of the confining matrix,
and break the Rouse assumption of time uncorrelation of the external forces acting on the tagged chain.

The model introduces a mixture A/B of bead-spring chains
with the same number of monomers, $N$. The monomer mass is $m = 1$.  
Monomers within a same chain are identical (of the same species A or B).
We simulate systems with $N =$ 4, 7, 10, 15, and 21. 
The monomer-monomer interaction potential is
$V_{\alpha\beta}(r) = 4\epsilon[(\sigma_{\alpha\beta}/r)^{12} - 7c^{-12} + 6c^{-14}(r/\sigma_{\alpha\beta})^{2}]$,
where $\epsilon=1$, $c = 1.15$ and $\alpha$, $\beta$ $\in$ \{A, B\}.
Potential and forces are continuous at the cutoff $r _{\rm c} = c\sigma_{\alpha\beta}$.
The interaction diameters are $\sigma_{\rm BB} =1$, $\sigma_{\rm AA} = 1.6\sigma_{\rm BB}$, 
and $\sigma_{\rm AB}=1.3\sigma_{\rm BB}$.
Chain connectivity is introduced by a FENE potential \cite{kremer,bennemannrouse},  
$V^{\rm FENE}_{\alpha\alpha}(r) = -kR_0^2 \epsilon\ln[ 1-(R_0\sigma_{\alpha\alpha})^{-2}r^2 ]$,
between consecutive monomers, with $k=15$ and $R_0 = 1.5$. 
The blend composition is  $x_{\rm B} = N_{\rm B}/(N_{\rm A} + N_{\rm B})$,
with $N_{\alpha}$ the number of $\alpha$-monomers. The total number of monomers is typically 2000-3000.
We use $ x_{\rm B} = 0.3$ and a packing fraction  $\phi = 0.53$ \cite{blendpaper}.
Temperature $T$, distance, 
and time $t$  are measured, respectively, in units of $\epsilon/k_B$, $\sigma_{\rm BB}$,
and $\sigma_{\rm BB}(m/\epsilon)^{1/2}$. 
Equilibrium simulations are performed in the microcanonical ensemble,
with 20-40 independent runs for statistical averages. The longest production runs
extend to $t \approx 4 \times 10^6$, corresponding to  
about 800 million time steps.

Monomer size disparity induces strong dynamic asymmetry between the
two components by decreasing $T$ \cite{blendpaper}. We define the `dynamic asymmetry',
$\chi = \tau_{\rm AA}/\tau_{\rm BB}$, 
with $\tau_{\alpha\alpha}$ the relaxation time
of the normalized density-density correlator $F_{\alpha\alpha}(q,t)$ 
for $\alpha$-$\alpha$ pairs \cite{blendpaper},
evaluated at the wavevector $q$ for the maximum of the static structure factor \cite{blendpaper}.
Thus, $\tau_{\alpha\alpha}$ is in the time scale of the structural $\alpha$-relaxation.
By using the definition $F_{\alpha\alpha}(q,\tau_{\alpha\alpha}) = 0.3$, $\chi$ increases,
from about 4 at $T = 1.5$, more than three decades by decreasing $T$.

\begin{figure}[t]
\begin{center}
\includegraphics[width=0.65\linewidth]{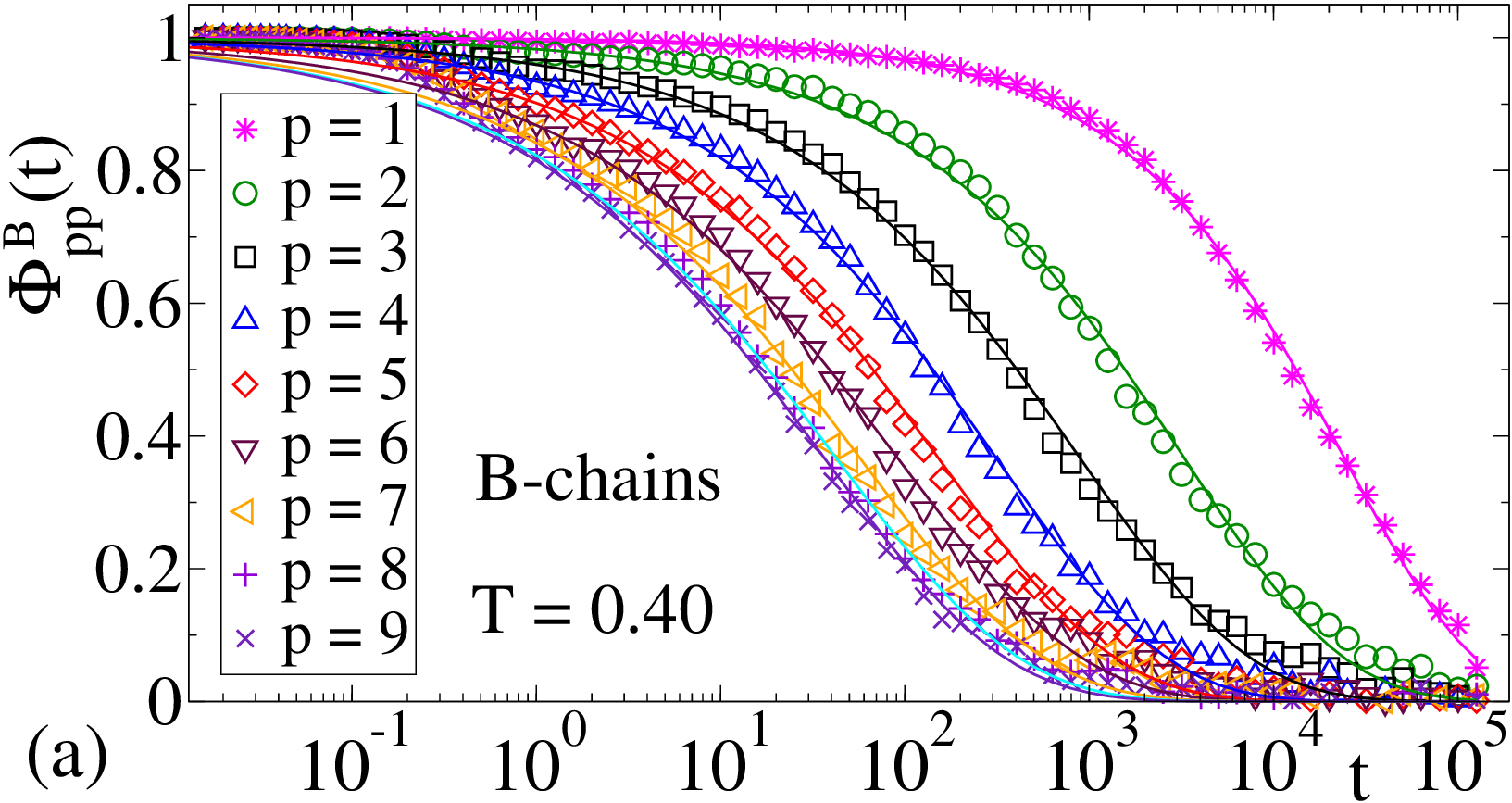}
\end{center}
\begin{center}
\includegraphics[width=0.76\linewidth]{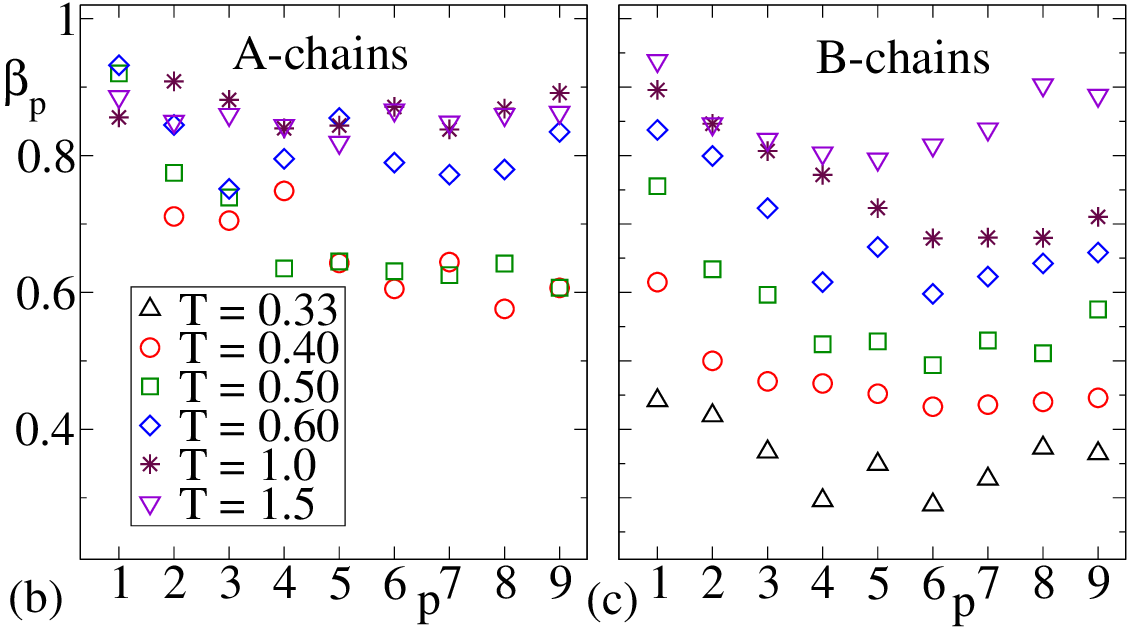}
\end{center}
\vspace{-5 mm}
\caption{Panel (a): Symbols are correlators of the Rouse modes for the B-chains, for $N = 10$
at $T = 0.40$. Lines are fits to a KWW function. Panels (b) and (c): For respectively
the A- and B-chains, $p$-dependence of the stretching exponents,
for $N =10$ at different temperatures. Symbol codes in (b) and (c) are identical.}
\label{fig:kwwrouse}
\end{figure}

Now we summarize the main predictions of the Rouse model \cite{doibook}.
A tagged chain is represented as 
a string of beads of equal mass connected by harmonic springs
of constant $3k_{\rm B}T/b^2$, with $b$ the bond length.
The effective interaction experienced by the chain is 
given by a friction coefficient $\zeta$ and a set of stochastic forces ${\bf f}_{j}$.
Excluded volume interactions are neglected.
The chain normal (Rouse) modes of index $p = 1,...N-1$ 
are defined as ${\bf X}_p (t) = N^{-1}\sum_{j=1}^{N}{\bf r}_j (t) \cos [jp\pi/N]$, and follow
the equations of motion
$2N\zeta {\bf\dot{X}}_p =  -k_p {\bf X}_p + {\bf g}_p $,
with $k_p = 24Nk_{\rm B}T b^{-2} \sin ^{2}[ p\pi/2N]$.
The external force for the $p$th-mode is
${\bf g}_p (t) = 2\sum_{j=1}^{N}{\bf f}_j (t) \cos[jp\pi/N]$.
Integration of the equations of motion yields
%
$\langle {\bf X}_p (t) \cdot {\bf X}_q (0)\rangle =
(2N\zeta ) ^{-2}\int^{t}_{-\infty} dt' e^{-(t-t')/\tau_p} 
\int^{0}_{-\infty} dt'' e^{t''/\tau_q}\langle {\bf g}_p (t') {\bf g}_q (t'')\rangle$
%
for the correlators of the Rouse modes.
The relaxation times
are given by
$\tau_p = (b^2 \zeta /12k_{\rm B}T)\sin^{-2}[p\pi/2N]$.
At this point no assumption has been made about the nature of the
forces ${\bf f}_{j}$. 
The Rouse model fully neglects spatial and time correlation
of the stochastic forces, i.e., 
$\langle {\bf g}_p (t') {\bf g}_q (t'')\rangle = 12N\zeta k_{\rm B}T\delta_{pq}\delta(t'-t'')$.
The former two approximations yield respectively orthogonality and
exponential relaxation of the Rouse correlators, leading to the expression
$\langle {\bf X}_p (t) \cdot {\bf X}_q (0)\rangle = 
(b^2 /8N)\sin^{-2}[p\pi/2N]\delta_{pq} \exp[-t/\tau_p]$.
Orthogonality and exponentiality of the Rouse modes are the two main predictions of the Rouse model,
and are the basis for the derivation of the correlators
probing chain relaxation \cite{doibook}.
By using the mentioned approximations, one obtains
$D_{\rm CM} = (N\zeta)^{-1}k_{\rm B}T$ 
for the diffusivity of the chain center-of-mass \cite{doibook}.
By introducing this result in $\tau_p$ (see above) we find 
the scaling relation for the Rouse relaxation times:
\begin{equation}
12N D_{\rm CM} \tau_p = b^2 \sin^{-2}[p\pi/2N].
\label{eq:rousescal}
\end{equation}

In the following we test the former predictions in the 
bead-spring blend here investigated. We compute normalized Rouse correlators for the $\alpha$-chains, 
$\Phi_{pq}^{\alpha}(t) = \langle {\bf X}^{\alpha}_p (t) \cdot {\bf X}^{\alpha}_q (0)\rangle /\langle [X^{\alpha}_p (0)]^2\rangle$.
According to orthogonality  $\Phi^{\alpha}_{pq}(0) = \delta_{pq}$. 
For both the A- and B-chains simulation data fulfill $| \Phi_{pq}^{\alpha} (0)| < 0.1$ 
for all the cases $p \neq q$ at all the temperatures, indicating than deviations from orthogonality are small.
Thus, we conclude that
there are only weak spatial
correlations between the external stochastic forces for the chains of both components,
and Rouse modes as defined above are good normal modes. This prediction is not affected by blending,
and in particular by the observed confinement effects on the fast component.
Now we test the exponentiality of the Rouse modes. 
Fig. \ref{fig:kwwrouse}a shows results
for normalized Rouse correlators of the B-chains for $N = 10$ at $T = 0.40$.
The decay of $\Phi_{pp}(t)$ from the plateau has been fitted \cite{notefit} to an empirical
Kohlrausch-Williams-Watts (KWW) function, $\propto \exp[-(t/\tau^{\rm K}_p)^{\beta_p}]$.
Figs. \ref{fig:kwwrouse}b and  \ref{fig:kwwrouse}c show, for respectively
the A- and B-chains, the $T$- and $p$-dependence of the stretching exponent $\beta_p$. 
According to the Rouse model relaxation of the Rouse correlators 
is purely exponential ($\beta_p = 1$). Clear deviations from this prediction are
observed by decreasing $T$ (i.e., by increasing $\chi$). 
At a same $T$, non-exponentiality is systematically more pronounced
for the B-chains, which show $\beta_p$-values of even 0.3 at $T = 0.33$.
Only at high temperature $\Phi_{pp}(t)$ is approximately exponential 
($\beta_p \gtrsim 0.8$) for both components.
These features are rather different from observations in the homopolymer case \cite{bennemannrouse},
which even at very low $T$ shows exponents $\beta_p \gtrsim 0.8$. Hence, blending induces
much stronger non-exponentiality for Rouse correlators.

\begin{figure}[t]
\begin{center}
\includegraphics[width=0.65\linewidth]{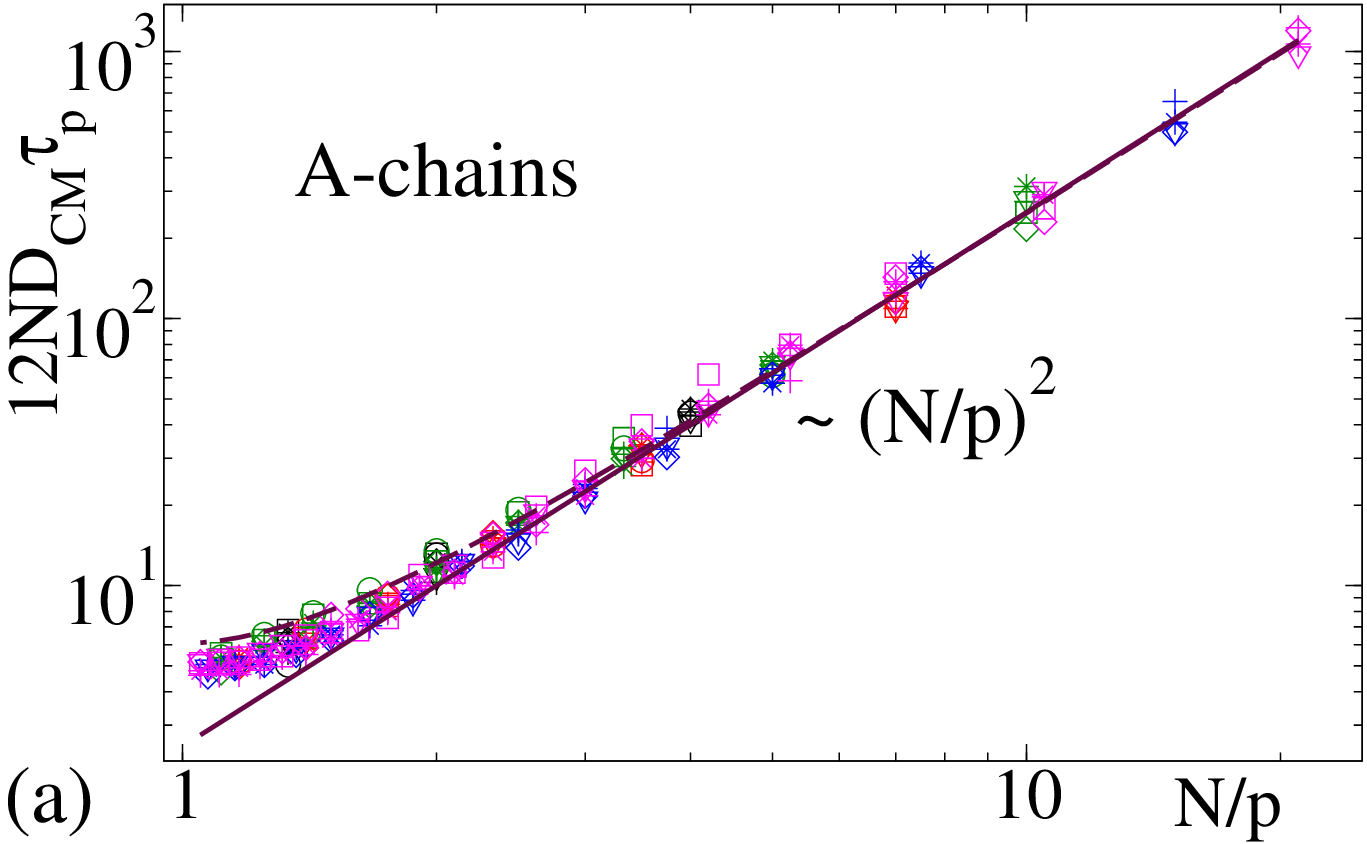}
\end{center}
\begin{center}
\includegraphics[width=0.62\linewidth]{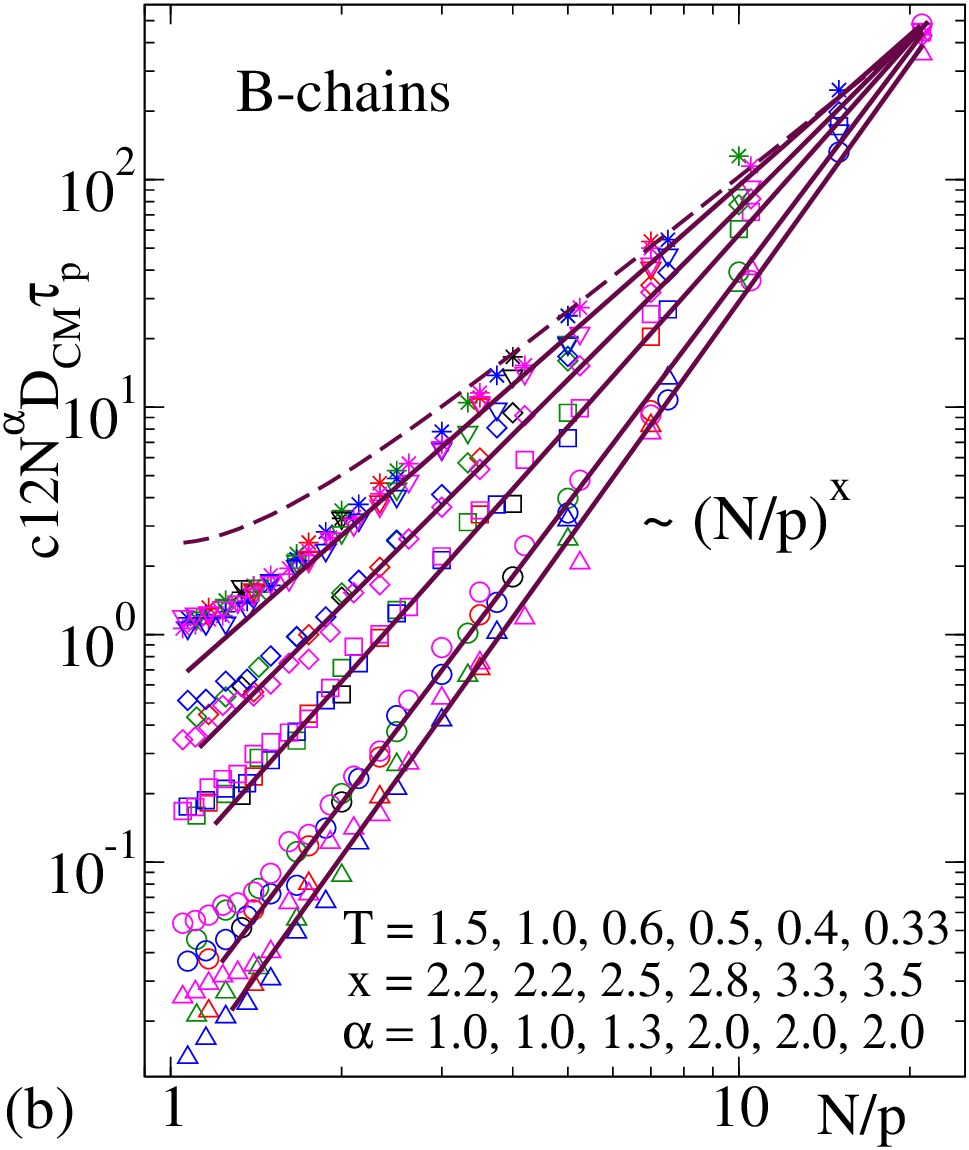}
\end{center}
\begin{center}
\includegraphics[width=0.64\linewidth]{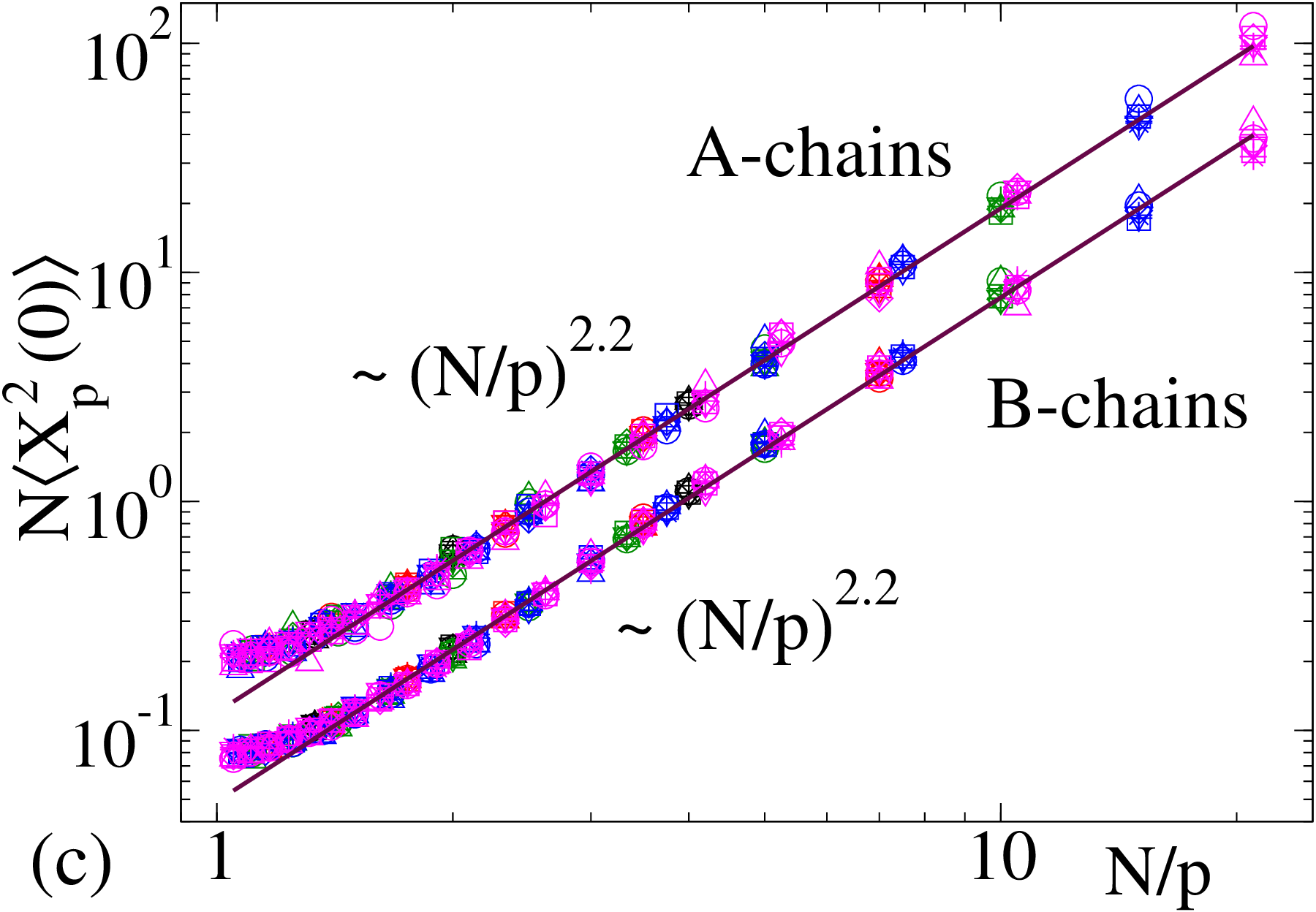}
\end{center}
\vspace{-5 mm}
\caption{Panels (a) and (b): Test of Rouse scaling for respectively the
A- and B-chains.  Identical symbols correspond to a same
$T$ (triangles: $T = 0.33$; circles: $T = 0.4$;
squares: $T= 0.5$; diamonds: $T = 0.6$; crosses: $T= 0.75$; stars: $T = 1.0$; 
down triangles: $T = 1.5$). Identical colors correspond to a same $N$ 
(black: $N = 4$; red: $N = 7$; green: $N = 10$; blue: $N = 15$; magenta: $N = 21$).  
Solid lines indicate power-law behavior $\sim (N/p)^{x}$. Dashed lines 
correspond to Eq. (\ref{eq:rousescal}) corrected by a shift factor
(see explanation in \cite{bennemannrouse}). For clarity of presentation, data in panel (b) are rescaled
by a $T$-dependent factor $c$. The latter drops about two decades 
from $c \lesssim 1$ at $T = 1.5$.
Panel (c): Static Rouse modes (scaled by $N$) for A- and B-chains.
Lines indicate an effective power-law $\sim (N/p)^{2.2}$. Color and symbol codes 
are as in (a) and (b).}
\label{fig:rousescal}
\end{figure}

In principle, non-exponentiality may be related
to a distribution of intrinsically exponential processes
originating from structural and/or dynamic heterogeneity. It is well-known that
a KWW function can be formally expressed as a sum of pure exponential functions
weighted by an adequate distribution $G$, though the latter does not necessarily have physical meaning.
Hence, we express the KWW function for the $p$th-mode as
$\exp[-(t/\tau^{\rm K}_p)^{\beta_p}] = \int d\tau _p G(\tau _p)\exp[-t/\tau _p]$,
where the distributed  values $\tau _p$  are the relaxation times of the different
exponential processes. Now we show that the latter is a good approximation
for the A-chains but unphysical for the B-ones. In other words, the Rouse modes
are basically exponential for the A-chains and
strongly non-exponential for the B-ones. Since the relaxation times for the elementary processes
would be  $\tau_p = \xi^{-1} \sin^{-2}(p\pi/2N)$ (see above), with $\xi$ a $p$-independent factor,
we can write
$\exp[-(t/\tau^{\rm K}_p)^{\beta_p}] = \int d\xi G(\xi)\exp[-\xi t \sin^{2}(p\pi/2N)]$,
with $\tilde{t} = t\sin^{2}(p\pi/2N)$ the `conjugated variable' of $\xi$.
By defining $\tilde{\tau} = \tau_{p}^{\rm K} \sin^{2}(p\pi/2N)$, 
we find $\exp[-( \tilde{t}/ \tilde{\tau} )^{\beta_p}] = \int d\xi G(\xi)\exp[- \tilde{t}\xi]$.
Since the right side of this equation is $p$-independent,
consistency requires the scaling condition
$\tau_{p}^{\rm K} \propto \sin^{-2}(p\pi/2N)$.
Fig. \ref{fig:rousescal} shows a test of Eq. (\ref{eq:rousescal}) 
--- also in the long wavelength limit $\tau_p \sim (N/p)^2$ --- for both components
at the investigated values of $N$ and $T$. 
This comparison also provides a test of the mentioned scaling condition \cite{notetau}. 
In agreement with observations for the non-entangled homopolymer case in
similar bead-spring models \cite{bennemannrouse} or in lattice models \cite{kreer}, 
data for the A-chains exhibit only small deviations, at small $N/p$.
Therefore, exponentiality of the Rouse modes for the A-chains is actually a good approximation,
and the observed non-exponentiality of the Rouse correlators basically originates from a distribution
of elementary exponential processes. The apparently stronger distribution effects (stronger stretching),
as compared to the homopolymer \cite{bennemannrouse}, might be related to structural and/or dynamic
heterogeneities induced by blending.

\begin{figure}[t]
\includegraphics[width=0.69\linewidth]{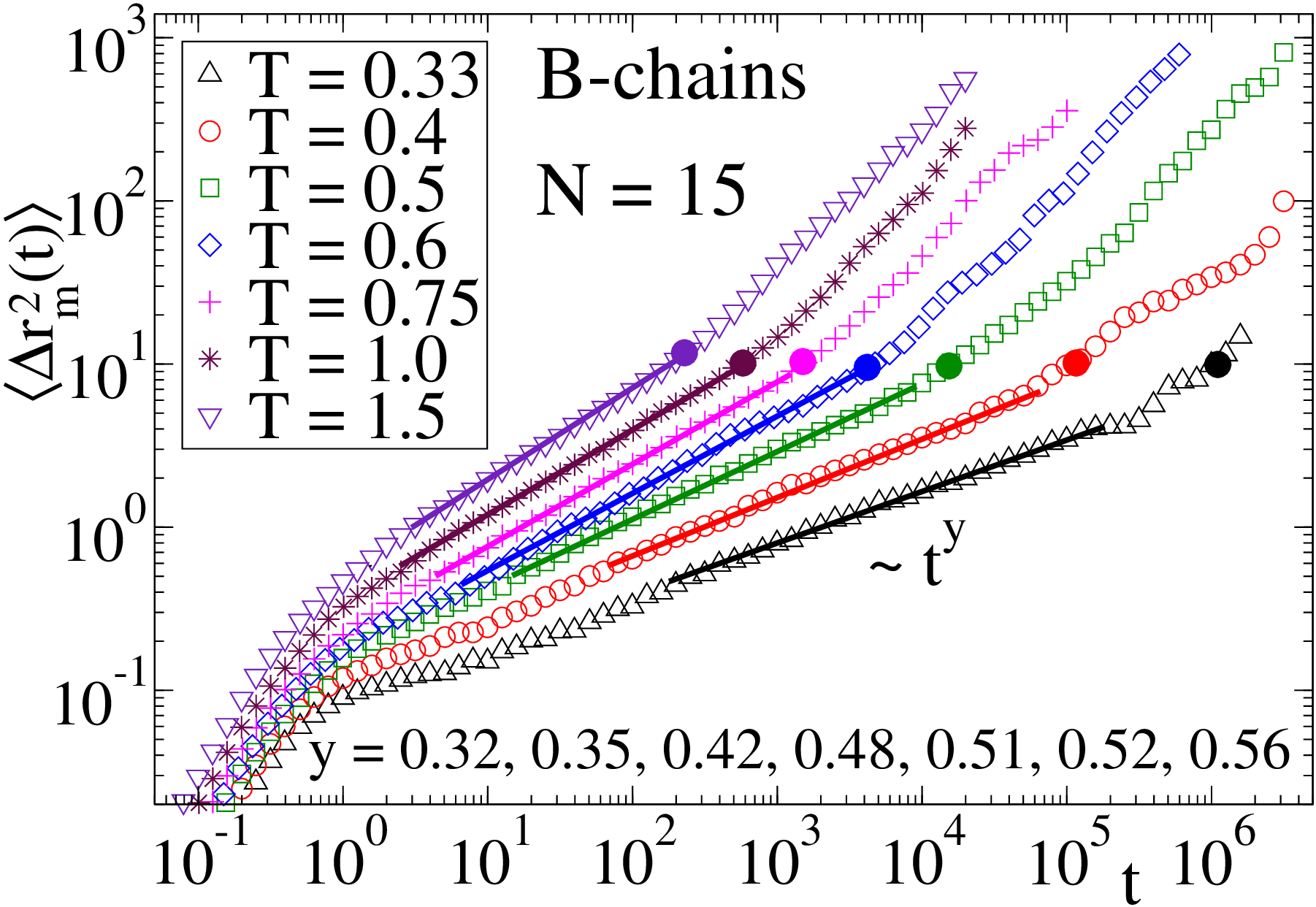}
\vspace{-2 mm}
\caption{Mean squared displacement of middle monomers of the B-chains, for $N = 15$.
Lines indicate an effective power-law $\sim t^y$. Filled circles mark the Rouse times $\tau_1$.}
\label{fig:msd}
\end{figure}

The B-chains exhibit a clear breakdown of Rouse scaling (Fig. \ref{fig:rousescal}b). 
This result demonstrates that for the fast B-component, 
though distribution efects are (as for the slow A-component) presumably present, the Rouse modes are
intrinsically, at low $T$, of strongly non-exponential nature. Eq. (\ref{eq:rousescal}) is approached
only in the high $T$ limit, where dynamic asymmetry ($\chi \approx 4$ at $T = 1.5$) and confinement
effects vanish. By decreasing $T$ we find an ultimate
regime $\tau_p  \sim (N/p)^{x \approx 3.5}$. This crossover is concomitant with the onset of
non-exponentiality of the Rouse modes. It must be stressed that this anomalous scaling behavior is
not related to particular static features of the Rouse modes.
Fig. \ref{fig:rousescal}c shows, for all the investigated values of $T$ and $N$, results
for the static Rouse modes, $\langle [X_p (0)]^2\rangle$. The latter display scaling behavior $\sim (N/p)^{2.2}$,
both for the A- and B-chains, in full analogy with observations for the homopolymer 
case \cite{bennemannrouse, notestatic}.  Therefore, anomalous scaling in Fig. \ref{fig:rousescal}b
is definitively a non-trivial dynamic effect.

Exponentiality of the Rouse modes originates
from the assumption of time uncorrelation of the stochastic forces (see above).
Results reported here evidence that the latter clearly breaks down for the fast component.
Several theoretical approaches based on projector operator techniques
incorporate density fluctuations around the tagged chain within
a memory kernel in a generalized Langevin equation (GLE).
Slow relaxation of the memory kernel induces strong non-exponentiality
of the Rouse modes \cite{kimmich}.  
The Rouse model (pure exponentiality) is recovered as a particular
case for which density fluctuations
around the tagged chain relax in a microscopic time scale \cite{kimmich}. This rough approximation
seems to work reasonably for simple non-entangled homopolymers and for the slow component
in the blend model here investigated, but it is clearly unrealistic
for the fast component at low $T$, which consists of chains diluted in a much slower host matrix.

The mentioned GLE methods, through approximations of the memory kernel within renormalized Rouse models,
predict for $\tau_p$ a crossover $(N/p)^{2} \rightarrow (N/p) ^{x \lesssim 3.5}$ in entangled homopolymers \cite{kimmich}.
The latter is in agreement with simulations \cite{shaffer,kremerprl,briels,richter} of homopolymers 
at fixed $T$ {\it by increasing the chain length} beyond the entanglement value $N_{\rm e} \sim 35$.
A similar crossover is observed in the non-entangled blend model here investigated {\it by increasing the dynamic asymmetry}
between the two components \cite{noterep}.
A further analogy in the fast B-component with predictions of GLE methods
for entangled homopolymers \cite{kimmich} 
is the observed anomalous scaling $\sim t^{y \approx 0.3}$ for the mean-squared displacement
in the time interval prior to the Rouse time $\tau_1$. Fig. \ref{fig:msd} shows for $N = 15$ a crossover,
concomitant with that of Fig. \ref{fig:rousescal}b, from Rouse-like scaling ($\sim t^{y \approx 0.5}$)
to an ultimate anomalous regime as the former.
Similar results are obtained for the B-chains at the other investigated chain lengths. 
On the contrary, a $T$-independent exponent $y \approx 0.6$  is obtained for the slow A-component \cite{blendpaper},
again in full analogy with the homopolymer case \cite{bennemannrouse}.

Within GLE methods observation of the mentioned anomalous power laws is directly
connected to slow relaxation of density fluctuations around the tagged chain \cite{kimmich}.
The latter may be induced by entanglement, but data reported here for the fast component
suggest that this is not a necessary ingredient. Analogies with entangled-like dynamics are indeed 
observed even for $N = 4$, provided that dynamic asymmetry in the blend is sufficiently strong. 
In summary, results presented here suggest a more general frame for chain
relaxation features usually associated to entanglement effects. They also open new
possibilities for the application of GLE methods in complex polymer mixtures.


We acknowledge support from NMP3-CT-2004-502235 (SoftComp, EU), 
MAT2007-63681 (Spain), and IT-436-07 (GV, Spain).


\begin{thebibliography}{9}
\bibitem{notestatic} Deviations from pure Rouse scaling, $\sim (N/p)^2$, for the static modes
are related to the breakdown of Gaussian 
chain statistics at small length scales \cite{bennemannrouse}.
\bibitem{noterep} Though confinement of the fast component in the blend may invoke the
common tube picture and associated reptation dynamics 
(data at low $T$ are even compatible with 
$D_{\rm CM} \propto N^{-2}$) \cite{doibook},
it was already stressed for the entangled homopolymer \cite{shaffer} that reptation
predicts a crossover in the $N$- but not in the $p$-scaling \cite{doibook}.
\end{thebibliography}
\end{document}